# Doping Dependence of Upper Critical Field of High-$T_c$ Cuprate Bi$_{2+x}$Sr$_{2-x}$CaCu$_2$O$_{8+\delta}$ Estimated from Irreversibility Field at Zero Temperature

Junichiro Kato[1,2], Shigeyuki Ishida[2]*, Tatsunori Okada[3], Shungo Nakagawa[2,4], Yutaro Mino[1,2], Yoichi Higashi[2], Takanari Kashiwagi[4], Satoshi Awaji[3], Akira Iyo[2], Hiraku Ogino[2], Yasunori Mawatari[2], Nao Takeshita[2], Yoshiyuki Yoshida[2], Hiroshi Eisaki[2], and Taichiro Nishio[1]*

[1]Department of Physics, Tokyo University of Science, Shinjuku, Tokyo 162-8601, Japan

[2]Research Institute for Advanced Electronics and Photonics, National Institute of Advanced Industrial and Science Technology (AIST), Tsukuba, Ibaraki 305-8568, Japan

[3]Institute for Materials Research, Tohoku University, Sendai 980-8577, Japan

[4]Graduate School of Pure & Applied Sciences, University of Tsukuba, Tsukuba, Ibaraki 305-8577, Japan

We investigated the temperature ($T$) dependence of the irreversibility field $H_{irr}(T)$ in high-critical-temperature cuprate Bi$_{2+x}$Sr$_{2-x}$Ca$_{1-y}$Y$_y$Cu$_2$O$_{8+\delta}$ (Bi-2212) single crystals over a wide range of hole doping level ($p$). $H_{irr}(T)$ was evaluated by measuring the magnetization hysteresis loop. The value of $H_{irr}(T)$ extrapolated to $T$ = 0 K [$H_{irr}(0)$], is either equal to or sets the lower boundary for the upper critical field at $T$ = 0 K [$H_{c2}(0)$]. $T_c$ shows a parabolic $p$-dependence (peak at $p$ = 0.16), whereas $\mu_0 H_{irr}(0)$ increases monotonically with $p$ by approximately one order of magnitude, from 19 T for the most underdoped sample ($p$ = 0.065, $T_c$ = 24 K) to 209 T for the most overdoped sample ($p$ = 0.200, $T_c$ = 75 K). The present results qualitatively agree with $H_{c2}(0)$ values evaluated from the specific heat measurements. The observed $p$-dependence of $H_{irr}(0)$ in Bi-2212 is distinct from those in YBa$_2$Cu$_3$O$_{7-\delta}$ and HgBa$_2$CuO$_{6+\delta}$, in which a pronounced dip structure appears in the underdoped region. Considering that the dip structures observed in these two systems are likely associated with the formation of competing orders (most likely field-induced charge orders), the present results indicate that the influence of the competing order in Bi-2212 is less prominent than that in the other two systems.

## 1. Introduction

In type-II superconductors, the upper critical field $H_{c2}$, which is the maximum magnetic field ($H$) up to which the superconducting state can be maintained, is one of the most fundamental parameters along with the critical temperature $T_c$. Below $H_{c2}$ and above the lower critical field $H_{c1}$, vortices appear in a superconductor, and typically, the vortices immediately solidify and form a lattice (the so-called vortex solid) [Fig. 1(a)]. As the vortices do not move freely, a zero-resistance state is realized.[1]

In the case of high-$T_c$ cuprate superconductors, $\mu_0 H_{c2}$ ($\mu_0$ is the vacuum magnetic permeability) is extremely high, exceeding 100 T in some cases, making it extremely difficult to determine its value through direct measurements. Furthermore, it is recognized that strong thermal fluctuations due to the two-dimensional layered crystal structure significantly alter the vortex phase diagram. As shown in Fig. 1(b), the vortex liquid phase, defined as a superconducting state in which the vortices move freely, extends within the region below $H_{c2}$.[2] In the vortex liquid phase, finite resistance is caused by the dissipative motion of the vortices. The boundary between the vortex solid and liquid phases is often denoted as the melting field $H_m$. While $H_m(T)$ is much lower than $H_{c2}(T)$ at $0 \ll T < T_c$, it is expected that $H_m(T)$ approaches $H_{c2}(T)$ as $T$ approaches 0 K (i.e., $H_m(0) \sim H_{c2}(0)$) because the effect of thermal fluctuation disappears at $T = 0$. From this figure, one can estimate $H_{c2}(0)$ by measuring $H_m(T)$ down to $T = 0$.

Another proposal for a possible vortex phase diagram highlights strong superconducting fluctuations in addition to thermal fluctuations. As shown in Fig. 1(c), the vortex liquid phase extends into the $T > T_c$ and $H > H_m(0)$ regions.[3] In this case, the value of $H_m(0)$ sets the lower boundary of $H_{c2}(0)$. To the best of our knowledge, it has not yet been determined whether Fig. 1(b) or Fig. 1(c) is correct.

The superconducting parameters of cuprate superconductors vary widely among different materials. They also depend significantly on the hole doping level ($p$). As for the determination of $H_{c2}$, various studies have been reported so far. Except for the cases of extremely underdoped (UD) or overdoped (OD) samples, where $T_c$ is very low, in most experiments, the transport properties were measured using pulsed high magnetic fields. For example, Ando et al.[4] evaluated $H_{c2}$ as the onset of the drop in electrical resistance and $H_m$ as zero resistance, through the interlayer resistivity ($\rho_c$) measurements using a 61 T pulsed field. For optimally doped (OP) (La,Sr)$_2$CuO$_4$ (La-214, $T_c \sim 35$ K) and Bi$_{2+x}$Sr$_{2-x}$CaCu$_2$O$_{8+\delta}$ (Bi-2212, $T_c \sim 80$ K), it was indicated that $\mu_0 H_{c2} \sim 60$ T and $\mu_0 H_m \sim 40$ T, respectively, at 4.2 K, with $H_m$ approaching $H_{c2}$ as $T$ approaches 0 K. Shibauchi et al.[5] measured $\rho_c$ of Bi-2212 up to 60 T and evaluated the field $H_{sc}(T)$ where a peak appears in the $H$ dependence of $\rho_c$ as a lower bound for $H_{c2}$. By extrapolating the $T$ dependence to 0 K, $\mu_0 H_{sc}(0)$ was estimated to be $\sim 100$ T for UD 68 K, $\sim 120$ T for OP 90 K, and $\sim 100$ T for OD 67 K, which roughly follows the parabolic $p$-dependence of $T_c$. Ramshaw et al.[6] measured the in-plane resistivity ($\rho_{ab}$) of YBa$_2$Cu$_3$O$_{7-\delta}$ (Y-123) at various $p$ values in a 60 T pulsed field and evaluated $H_m$ as the field where finite resistance appears. The extrapolated $\mu_0 H_m(0)$ value is $\sim 120$ T for the OP sample. They also reported that $H_m(0)$ exhibited characteristic $p$ dependence, with a minimum near $p = 1/8$. The same group later demonstrated that $H_{c2}(0)$ determined from thermal conductivity measurements in a 45 T steady field agreed with $H_m(0)$ [denoted as $H_{vs}(0)$ in their paper].[7] They showed that $H_{c2}(0)$ has two peaks at $p \sim 0.08$ and 0.19 with $\mu_0 H_{c2}(0) \sim 50$ T and 150 T, respectively. It is argued that the minimum around $p = 1/8$ is due to the development of charge order, which competes with superconductivity. Chan et al.[8] measured the $\rho_{ab}$ and Hall resistances of HgBa$_2$CuO$_{6+\delta}$ (Hg-1201) using pulsed fields of up to 100 T to evaluate $H_m(0)$. Their results showed a non-monotonic $p$ dependence, with a minimum ($\sim 40$ T) around $p \sim 0.10$ and a maximum ($\sim 110$ T) at $p \sim 0.19$, similar to the behavior observed in Y-123. These results support the vortex phase diagram in Fig. 1(b).

Wang et al. proposed the vortex phase diagram shown in Fig. 1(c) based on the Nernst effect measurements on La-214, Bi-2212, and Bi$_2$(Sr,La)$_2$CuO$_{6+\delta}$ (Bi-2201). They defined $H_{c2}$ and $T_{onset}$ as the $H$ and $T$, respectively, at which the Nernst signals related to the vortices disappeared.[9,10] In their study, the Nernst signal was observed even at $H > H_m(0)$ and $T > T_c$. As for the $p$ dependence, $H_{c2}$ decreases with $p$, which is contrary to previous reports in which $H_{c2}$ increases with $p$. However, there are different opinions regarding the origin of the Nernst signal above $T_c$, and whether the vortices exist above $T_c$ has not been settled even at this time.[11]

Recently, studies on Y-123 which exhibits a developed competing order (charge order) have reported the existence of a "second vortex solid phase" at low $T$ and high $H$, through magnetic torque measurements in pulsed fields.[12] This second vortex solid phase extends to the region above $H_{c2}(0)$ determined from thermal conductivity measurements, supporting the vortex phase diagram in Fig. 1(c). Additionally, $\rho_{ab}$ measurements of Y-123,[13] La-214, and Bi-2201[14] reported the appearance of an "unconventional quantum vortex matter state" or "anomalous vortex liquid phase" in the low-$T$ region characterized by a non-ohmic behavior in the regions where competing charge orders exist.

In addition to transport measurements, Mittag et al.[15] and Dewhurst et al.[16] performed magnetization hysteresis loop (MHL) measurements on Bi-2212 single crystals. By measuring the MHLs, one can experimentally determine the irreversibility field $H_{irr}(T)$ at which the vortices depin and move freely, thus corresponding to $H_m(T)$. Using this method, Mittag et al.[15] and Dewhurst et al.[16] estimated $\mu_0 H_{irr}(0)$ to be approximately 120 and 147 T, respectively, for OP Bi-2212.

Moreover, Tallon et al. carried out systematic specific heat measurements on polycrystalline Bi-2212 samples with various $p$ values and calculated $H_{c2}$ by resolving the superconducting condensation energy. They showed that while $T_c$ exhibits a parabolic $p$ dependence with a peak at $p = 0.16$, $\mu_0 H_{c2}(0)$ exhibits a maximum value of ~100 T around $p \sim 0.19$, corresponding to the OD region.[17] They interpreted that the reduction in $H_{c2}$ in the UD region is due to the formation of a pseudogap, which results in the suppression of the superconducting condensation energy.

This review of previous studies shows that systematic evaluation of $H_{c2}$, comparison between different measurement methods, and comparison between different materials are still incomplete, and determining the magnitude and $p$-dependence of $H_{c2}$ in high-$T_c$ cuprate superconductors remains an important challenge. In this study, we performed systematic MHL measurements using high-quality Bi-2212 single crystals with precisely controlled $p$ values and evaluated the $p$-dependence of $H_{irr}(0)$ as a lower boundary of $H_{c2}(0)$. We found that $\mu_0 H_{irr}(0)$ increased monotonically from 19 T for the most UD ($p = 0.065$) to 209 T for the most OD ($p = 0.200$) samples. These results qualitatively agree with Tallon's estimates of $H_{c2}(0)$ from specific heat measurements.[17] Although the magnitude of $H_{irr}(0)$ is comparable to those in Y-123[6,7] and Hg-1201,[8] we did not observe any minima of $H_{irr}(0)$ at a specific $p$, suggesting no clearly visible effect from charge order development. This indicates that the electronic phase diagram of Bi-2212 differs from those of Y-123 and Hg-1201.

## 2. Methods

### 2.1 Sample preparation

Single-crystal Bi-2212 samples were grown using the traveling-solvent floating-zone (FZ) technique. To cover a wide $p$ range from the heavily UD region to the OD region, two types of single crystals were synthesized: nonsubstituted Bi-2212 and Y-substituted $Bi_2Sr_2Ca_{1-y}Y_yCu_2O_{8+\delta}$ (Bi(Y)-2212). For Bi-2212, one can tune $p$ from UD to OD by post-annealing as described below. In Bi(Y)-2212, the substitution of $Ca^{2+}$ sites with $Y^{3+}$ reduced $p$ of the as-grown samples, and post-annealing changed $p$ from heavily UD to slightly UD.

For Bi-2212, powders of $Bi_2O_3$, $SrCO_3$, $CaCO_3$, and CuO (all 99.99% purity) were mixed in a composition ratio of Bi:Sr:Ca:Cu = 2.15:1.85:1.00:2.00. For Bi(Y)-2212, powders of $Bi_2O_3$, $SrCO_3$, $CaCO_3$, $Y_2O_3$ (99.99% purity), and CuO were mixed in a composition ratio of Bi:Sr:Ca:Y:Cu = 2.00:2.00:0.80:0.20:2.00. Both samples underwent several cycles of calcination at 730–810 °C and intermediate mixing to obtain uniform polycrystalline powders. Premelting and crystal growth were conducted in an atmospheric environment using a FZ furnace with infrared radiation (halogen lamps) from Canon Machinery and Crystal Systems.

Four Bi-2212 samples and two Bi(Y)-2212 samples were cut from the obtained single-crystal rods using a razor blade. The dimensions (width ($w$) × length ($l$) × thickness ($t$)) of the samples were $0.748 \times 0.820 \times 0.034$ mm$^3$ (Bi-2212 #1), $0.783 \times 0.783 \times 0.035$ mm$^3$ (Bi-2212 #2), $0.787 \times 0.808 \times 0.033$ mm$^3$ (Bi-2212 #3), $0.716 \times 0.762 \times 0.038$ mm$^3$ (Bi-2212 #4), $0.785 \times 1.274 \times 0.073$ mm$^3$ (Bi(Y)-2212 #Y1), and $1.298 \times 1.370 \times 0.068$ mm$^3$ (Bi(Y)-2212 #Y2).

The chemical compositions of the single crystals were analyzed using scanning electron microscopy (SEM, Hitachi High-Technologies, TM3000) equipped with energy-dispersive X-ray spectroscopy (EDX, Oxford Instruments, SwiftED3000). The average composition ratio of Bi to Sr in the four Bi-2212 samples was 2.16:1.84, close to the nominal ratio. For the two Bi(Y)-2212 samples, the Bi to Sr ratio was similar (Bi:Sr = 2.04:1.96); however, the amount of Y substitution into Ca site slightly differed; Ca:Y = 0.53:0.47 for #Y1 and Ca:Y = 0.60:0.40 for #Y2. X-ray diffraction (XRD) measurements (Rigaku, Ultima VI) were performed on the $ab$ plane of the samples, which confirmed the absence of impurity phases and the 00$l$ peaks of narrow widths (e.g. the full width at half maximum of the 0020 peak observed at ~60 degrees was typically less than 0.1 degrees).

*2.2 Control of p by post-annealing*

Post-annealing was performed to control the oxygen content of the single crystals, and thus adjust $p$. Post-annealing was conducted in a tubular furnace under a controlled atmosphere of oxygen ($O_2$) or oxygen/nitrogen ($O_2/N_2$) gas flow for 1–7 d. To prevent unintended oxidation/reduction of the samples during the cooling process, which may cause inhomogeneous oxygen distributions, the samples were quenched from the annealing temperature ($T_a$) to room temperature within one minute under a gas flow atmosphere. The annealing conditions applied to the single-crystal samples are listed in Table I. Multiple conditions listed for a single sample indicate that the annealing and measurement processes were repeated for the same sample.

## 2.3 Determination of $T_c$ and estimation of $p$

$T_c$ was determined based on the $T$ dependence of the magnetization ($M$). $M$ was measured using a SQUID magnetometer (Quantum Design, MPMS-3) during the zero-field-cooling (ZFC) and field-cooling (FC) processes under $\mu_0 H$ of 0.3 mT applied along the $c$ axis. $T_c$ was defined as the intersection at which a linear fit of the $T$ dependence of the diamagnetic signal in the FC process was extrapolated to $M = 0$. To estimate $p$, we use the following empirical parabolic dependence:[18]

$$T_c = T_c^{OP}[1 - 82.6(p - 0.16)^2], \tag{1}$$

where $T_c^{OP}$ is the $T_c$ of the OP sample.

## 2.4 Evaluation of $H_{irr}(T)$ and estimation of $H_{irr}(0)$

The MHL measurements were performed with $H$ parallel to the $c$ axis using a 9T superconducting magnet (Quantum Design, PPMS) and 25T/20T cryogen-free superconducting magnets (25T/20T-CSM)[19,20] at the High Field Laboratory for Superconducting Materials, Tohoku University. The $M$ was measured under $H$-increasing and decreasing processes using a vibrating sample magnetometer (VSM) with resolution of ~5 × 10⁻⁹ A·m² (5 × 10⁻⁶ emu) for the PPMS and ~5 × 10⁻⁸ A·m² (5 × 10⁻⁵ emu) for 25T/20T-CSM. The $H$ sweep rate was ~5 mT·s⁻¹ both for the PPMS and the 25T/20T-CSM.

At high temperatures ($T$s), $H_{irr}(T)$ was determined directly from the MHL in the $H$ range used in this study. However, with decreasing $T$, $H_{irr}(T)$ increases and exceeds the measurable $H$ range. In such cases, we use the scaling law of the pinning force density $F_p$.[15,16]

Based on the Bean's critical state model,[21] $F_p$ is expressed as

$$F_p(H) = J_c(H) \times \mu_0 H, \tag{2}$$

where $J_c(H)$ denotes the critical current density. The $J_c(H)$ values were obtained from the MHL measurements by evaluating $\Delta M(H)$, which is the difference between $M(H)$ measured in $H$-increasing and $H$-decreasing processes.

$$J_c(H) = \frac{2\Delta M(H)}{w(1-w/3l)}, \tag{3}$$

where $w$ and $l$ ($w < l$) are the in-plane dimensions of the sample. The units are A·m⁻² for $J_c(H)$, A·m⁻¹ for $\Delta M(H)$, and m for $w$ and $l$.

Empirically, $F_p$ is a function of $T$ and $H$, and is expressed as[22,23]

$$F_p(H,T) = A[H_{irr}(T)]^m\, h^\alpha(1-h)^\beta, \tag{4}$$

where $h = H/H_{irr}$. Here $A$ is a constant and $\alpha$ and $\beta$ are parameters determined by the geometry of the pinning centers and by the interaction between vortices and pinning centers. At each $T$, Eq. (4) takes its maximum value $F_p^{max}(T)$ at the field $H_{max}$ satisfying $h_{max} = H_{max}/H_{irr} = \alpha/(\alpha+\beta)$. The normalized pinning force $f_p(h)$ is expressed as

$$f_p(h) = \frac{F_p(h,T)}{F_p^{max}(T)} = \frac{(\alpha+\beta)^{\alpha+\beta}}{\alpha^\alpha\beta^\beta}h^\alpha(1-h)^\beta, \tag{5.1}$$

or using a more experimentally-accessible parameter $h^* = H/H_{max}$,

$$f_p(h^*) = (\alpha/\beta)^\beta \cdot h^{*\alpha}[(\alpha+\beta)/\alpha - h^*]^\beta. \tag{5.2}$$

Note that Eq. (5.1) [Eq. (5.2)] depends only on $h$ ($h^*$) and not on $T$. This implies that the $F_p - H$ curves measured at different $T$s are scaled to a universal curve when the pinning mechanism does not change, i.e., $\alpha$ and $\beta$ are constant. Accordingly, once the $F_p - H$ relationship up to $H_{irr}$ is established using the data at several $T$s, one can estimate $H_{irr}$ at other $T$s, even if the $F_p - H$ curves only in the $H < H_{irr}$ region are known.

Once $H_{irr}(T)$ at each $T$ was determined, $H_{irr}(0)$ was estimated by fitting the data ($T_0$ is a fitting parameter) using the following equation:

$$H_{irr}(T) = H_{irr}(0)\exp\left(-\frac{T}{T_0}\right), \tag{6}$$

which is an empirical formula applicable to a variety of high-$T_c$ cuprate superconductors.[15,16]

Equation (4) also indicates that a power-law relationship exists between $F_p^{max}(T)$ and $H_{max}(T)$

as follows:

$$F_p{}^{max}(T) = A \frac{\alpha^\alpha \beta^\beta}{(\alpha+\beta)^{\alpha+\beta}} [H_{max}(T)]^m. \tag{7}$$

## 3. Results

### 3.1 Determination of $T_c$ and estimation of $p$

Figures 2(a)–2(c) show the $T$ dependence of $M$ measured during the FC process [$M_{FC}(T)$] for representative samples. The $M_{FC}$ values are normalized by the magnitude of those at the lowest $T$s [$|M_{FC}(T \ll T_c)|$]. In Fig. 2(a), the $M_{FC}(T)$ values for Bi-2212 samples annealed under $O_2$-flow conditions are plotted, covering the OP to OD regions. When $T_a$ was lowered from 800 °C (blue) to 400 °C (purple); that is, when a stronger oxidation condition was applied, the $T_c$ systematically decreased from 87.6 K to 76.5 K, confirming that OD samples were obtained. Figure 2(b) shows the $M_{FC}(T)$ of Bi-2212 annealed in an $O_2/N_2$ mixture, corresponding to the reduction condition. As $T_a$ was raised from 450 °C (green) to 630 °C (orange), that is, by applying stronger reduction, $T_c$ systematically decreased from 83.4 K to 69.7 K, indicative of UD samples (note that the sample with $T_c$ of 53.8 K (pink) was obtained after the additional reduction annealing of the sample with $T_c$ of 69.7 K (orange)). Finally, for the Bi(Y)-2212 samples, as $T_a$ increased from 400 °C to 800 °C, $T_c$ decreased from 79.6 K (purple) to 54.5 K (brown) for Y#2, and from 61.2 K (blue) to 35.6 K (red) for Y#1, respectively. These samples cover the UD region.

The broadness of the superconducting transition evaluated by the $T$ range, where $M_{FC}(T)$ decreases from 10% to 90% of the full transition, was smaller than 5 K for the slightly UD to OD Bi-2212 samples, indicating good homogeneity. For the UD Bi-2212, the transition broadened, reaching 27.7 K for the heavily UD sample with $T_c = 53$ K. Meanwhile, in the case of Bi(Y)-2212 samples, the transition width was around 10 K even for the heavily UD samples.

Then, $p$ was estimated from $T_c$ using Eq. (1), where $T_c{}^{OP}$ was taken as 87.6 K for Bi-2212 and 95 K for Bi(Y)-2212. The obtained relationship between $T_c$ and $p$ is plotted in Fig. 2(d), showing that Bi-2212 and Bi(Y)-2212 samples cover $p = 0.092$–$0.200$ and $p = 0.065$–$0.116$, respectively.

### 3.2 Evaluation of $H_{irr}(T)$ through MHLs

First, we explain the evaluation process of $H_{irr}(T)$ in detail by presenting the experimental results for Bi(Y)-2212 with $T_c$ of 61.2 K. The low $H_{irr}(T)$ of this sample allowed the direct measurement of $H_{irr}(T)$ down to (relatively) low $T$s. Figure 3 shows the MHLs measured at $T = $ 55−35 K (a), 30−20 K (b), and 20−4.2 K (c). Figures 3(d)−3(f) show enlarged views near $M = $ 0. The $M$ for $H$-increasing and $H$-decreasing exhibited different values, that is, hysteresis, and the size of the MHL increased with decreasing $T$. The solid vertical arrows in Figs. 3(d)−3(f) indicate $H_{irr}$ at each $T$, corresponding to the points where the MHL closes. Note that we here defined $H_{irr}$ as a point at which $J_c$ shown in Fig. 4(a) becomes less than $3 \times 10^6$ A·m$^{-2}$, while $H_{irr}$ depends on the $J_c$ criterion, as described later. As expected, $H_{irr}$ increases with decreasing $T$.

As shown in Figs. 3(c) and 3(f), below 10 K, $H_{irr}(T)$ exceeded the maximum $H$ of the measurement (9 T). To evaluate $H_{irr}(T)$ at low $T$s, we first calculated $J_c$ using Eq. (3). The $H$ dependence of $J_c$ [$J_c(H)$] at $T = 18$–4.2 K is shown in Fig. 4(a). The $H$ dependence of $F_p$ [$F_p(H)$] was then derived from $J_c(H)$ using Eq. (2). The obtained $F_p(H)$ at each $T$ are shown in Fig. 4(b). While $J_c$ monotonically decreases with $H$, $F_p$ shows a maximum ($F_p^{max}$) at $H_{max} << H_{irr}$ (inverse triangles indicate $F_p^{max}$ at each $T$). $F_p^{max}$ and $H_{max}$ increased towards low $T$.

Figure 4(c) shows the $h^*$ dependence of $f_p$ [$f_p(h^*)$], obtained by normalizing $F_p(H)$ with $F_p^{max}$ and $H_{max}$ at each $T$. According to Fig. 4(c), $f_p(h^*)$ below 14 K appears to collapse into a single curve, indicating that $f_p$ is uniquely determined by $h^*$; that is, scaling is fulfilled at low $T$s within the range of measurements. As $T$ increased, the data began to deviate from the universal curve, which was most prominent at $T = 18$ K (purple).

To further verify the validity of the scaling at low $T$s, $F_p^{max}(T)$ is plotted against $H_{max}(T)$ in Fig. 4(d) as a double-logarithmic graph. The $F_p^{max}(T)$ between 16 K to 4.2 K show a linear dependence as fitted by the dashed line, yielding $F_p^{max}(T) \propto H_{max}(T)^m$ with $m \sim 1.8$. This is consistent with Eq. (7), and the obtained $m$ is close to the reported value ($m \sim 2$) in the previous study.[15] In addition, the data at 18 K deviate from the dashed line, which corresponds to the deviation of $f_p(h^*)$ at 18 K from the universal curve shown in Fig. 4(c). These results also support the fact that scaling is fulfilled at low $T$ below 16 K.

This confirmed that $H_{irr}/H_{max}$ is constant in the low $T$ region, and $H_{irr}(T)$ can be estimated from $H_{max}(T)$. However, since $H_{irr}$ depends on the $J_c$ criterion, uncertainty arises in determining $H_{irr}/H_{max}$. In Fig. 4(e), the $T$-dependences of $H_{irr}/H_{max}$ values obtained using different $J_c$ criteria ($J_c = 1, 2, 3, 5, 7,$ and $10 \times 10^6$ A·m$^{-2}$) are plotted. The difference between $H_{irr}/H_{max}$ for different criteria decreases as $T$ decreases, and $H_{irr}/H_{max}$ values converge in the range of $5 \pm 0.5$ in low-$T$ region. Thus, $H_{irr}(T)$ and its uncertainty were evaluated using $H_{irr}/H_{max} = 5 \pm 0.5$. Note that the $H_{irr}/H_{max}$ value of 5 agrees with the values reported in previous studies for Bi-2212 and Bi-2201.[15,24]

Figure 5(a) shows $H_{irr}(T)$ plotted in a semi-logarithmic graph. The filled blue circles represent the data directly determined from the MHLs and the red diamonds represent the values estimated using the relationship $H_{irr}/H_{max} = 5$. Overall, $H_{irr}(T)$ exhibits an exponential decay from the lowest $T$ (4.2 K) to the highest $T$ near $T_c$. This exponential behavior suggests that the thermal fluctuations enhanced by the large anisotropy in Bi-2212 primarily determine $H_{irr}(T)$. More specifically, $H_{irr}(T)$ shows a kink-like feature at $T \sim 20$ K, or $\mu_0 H_{irr}(T) \sim 0.1$ T. Previous studies have pointed out that the three- to two-dimensional crossover of the vortex lattice occurs at $\mu_0 H \sim 0.1$ T in Bi-2212, and that the kink feature is associated with the crossover.[2,16,25] This behavior of $H_{irr}(T)$ is related to the fact that the scaling of $f_p(h^*)$ in Bi-2212 is only fulfilled in the low-$T$ region [Fig. 4(c)].

Figure 5(b) shows an enlarged view of the low-$T$ region. Assuming that the thermal fluctuations dominantly determine $H_{irr}(T)$ and that no additional vortex phase transition occurs at low $T$, $H_{irr}(0)$ can be estimated by fitting the data using Eq. (6). The fitting (the dashed line in Fig. 5(b), where the $T$ range from 4.2 to 13 K was used) yields the parameters $\mu_0 H_{irr}(0) = 87.5$ T, and $T_0 = 4.1$ K. Thus, we determined $\mu_0 H_{irr}(0)$ of Bi(Y)-2212 with $T_c = 61.2$ K ($p = 0.094$) to be $87.5 \pm 8.8$ T, taking into account the uncertainty of $H_{irr}/H_{max} = 5 \pm 0.5$.

### 3.3 p dependence of $H_{irr}(T)$

Figures 6(a1)–6(a10) show $J_c(H)$ in semi-log plots for the Bi-2212 and Bi(Y)-2212 samples with $p$ values from 0.073 (UD) to 0.200 (OD). The magnitude of $J_c$ varies significantly with $p$. For example, $J_c(0)$ at 10 K (blue) increases from $1 \times 10^9$ A·m$^{-2}$ ($p = 0.073$) to $2 \times 10^{10}$ A·m$^{-2}$ ($p = 0.200$). Also, the magnitude of $H_{irr}$ increases with $p$; $H_{irr}(10K)$ is smaller than 9 T up to $p = 0.094$, whereas it exceeds 25 T for $p \geq 0.160$. Figures 6(b1)–6(b10) show $F_p(H)$ calculated from $J_c(H)$. For all the samples, $F_p$ increased significantly with decreasing $T$, and the values of $H_{max}$ (inverse triangles in the figures) also increased with decreasing $T$. Figures 6(c1)–6(c10) show the $h^*$ dependence of $f_p$ [$f_p(h^*)$]. All $f_p(h^*)$ curves exhibited scaling behavior. It is commonly observed that $f_p$ becomes zero at $h^* \sim 5$ for all samples, indicating that the ratio of $H_{irr}$ to $H_{max}$ is approximately 5, which is common to all $p$ values. Therefore, we employed $H_{irr}/H_{max} = 5$ for all samples to estimate $H_{irr}(T)$ at low $T$s.

Figure 7(a) shows the semi-log plot of $H_{irr}(T)$ at $T < 16$ K for the samples evaluated in this study. For all samples, $H_{irr}(T)$ exhibited a linear $T$ dependence on a semi-log graph, indicating an exponential increase with decreasing $T$. Following the previous example, fitting was performed using Eq. (6); the results are shown as solid lines. The $H_{irr}(0)$ values obtained by fitting are plotted against $p$ in Fig. 7(b) together with $T_c$ (dashed curve). The error bars reflect the uncertainty in the $H_{irr}/H_{max}$ value. In contrast to the parabolic $p$-dependence of $T_c$, $H_{irr}(0)$ shows a monotonic increase from $p = 0.065$ to 0.200.

### 4. Discussion

*4.1 Comparison of $H_{c2}$ in Bi-2212 with previous studies*

The main outcomes of this study are twofold: (i) determination of the magnitude of $H_{irr}(0)$ in Bi-2212, and (ii) elucidation of its $p$ dependence. $H_{irr}(0)$ increases monotonically and smoothly with $p$ in the present $p$ range. For the most UD samples ($p = 0.065$), $\mu_0 H_{irr}(0)$ was 19 T. $\mu_0 H_{irr}(0)$ increased to 173 T for the OP sample ($p = 0.160$), where $T_c$ was the highest. With a further increase in $p$, $T_c$ began to decrease, whereas $\mu_0 H_{irr}(0)$ continued to increase, reaching 209 T in the most OD samples ($p = 0.200$). Thus, $H_{irr}(0)$ increased by a factor of 11, from $p = 0.065$ to 0.200.

As mentioned previously, the $H_{irr}(0)$ value obtained in this measurement defines the lower limit of $H_{c2}(0)$. Accordingly, one can estimate the upper limit of the coherence length $\xi$ using the formula $\mu_0 H_{c2} = \phi_0/2\pi\xi_{ab}^2$ for $H /\!/ c$ where $\phi_0$ is flux quantum. The values of $\xi_{ab}$ are estimated as 4.2 nm for $p = 0.065$, 1.4 nm for $p = 0.160$, and 1.3 nm for $p = 0.200$. In the OD sample, $\xi_{ab}$ is at most 3.4 times the in-plane Cu-Cu distance (0.38 nm).

Here, we compare our results with those of previous studies. Mittag et al.[15)] and Dewhurst et al.[16)] performed MHL measurements on OP Bi-2212 single crystals. They estimated $\mu_0 H_{irr}(0)$ to be 120 T and 147 T, respectively, which is in good agreement with the present results.

Next, according to a report by Tallon et al.,[17)] $H_{c2}(0)$, which was derived from the analysis of the specific heat data of polycrystalline Bi-2212, increases with $p$, reaching a maximum at $p = 0.19$, and then begins to decrease. In Fig. 8, we compare these results (red diamonds) with those of our study (blue circles). The general trends, both in magnitude and $p$-dependence, were in good agreement with each other. If one looks closer, while $H_{irr}(0)$ evaluated in this work increases with $p$ up to 0.200, $H_{c2}(0)$ evaluated from specific heat measurement decreases in this $p$ range. Moreover, the specific heat results are approximately half of those from the MHL experiments. The results of the specific heat measurements are plotted against the right y-axis, whose scale is half that of the left axis. It should be noted that the physical parameter obtained from MHL is $H_{irr}(0)$, whereas that obtained from the specific heat measurement is $H_{c2}(0)$. The present results, if both are correct, imply $H_{irr}(0) > H_{c2}(0)$, which is apparently incompatible with the vortex phase diagrams depicted in Figs. 1(b) and 1(c) in which $H_{c2}(0) \geq H_{irr}(0)$ is required. This incompatibility suggests that $H_{c2}(0)$ determined indirectly from the specific heat data was underestimated, although overestimation of $H_{irr}(0)$ from MHL in this study remains possible.

Regarding the origin of the $p$ dependence, Tallon et al. proposed that the remarkable decrease in $H_{c2}(0)$ in the UD region is due to the formation of a pseudogap, which competes with superconductivity and reduces the superconducting condensation energy. The $p$-dependence of $H_{irr}(0)$ observed in this study can be understood similarly. For the OD region ($p > 0.19$), differences in the chemical composition (Pb-substituted Bi-2212 samples were used for the specific heat measurements) possibly resulted in the different behaviors. To resolve this issue, $H_{irr}(0)$ of Pb-substituted samples should be investigated.

In Fig. 8, we also plot the values of $H_{c2}(T_c)$ (green squares) estimated by Wang et al.[9] through Nernst-effect measurements of Bi-2212 single crystals. The $H_{c2}$ values shown here are for $T = T_c$ and not for $T = 0$. However, in their picture where $T_c << T_{onset}$ [Fig. 1(c)], the magnitude of $H_{c2}$ does not change significantly from $T = T_c$ to 0, thus $H_{c2}(T_c) \sim H_{c2}(0)$ is expected. According to their results, $\mu_0 H_{c2}(T_c)$ is estimated to be 140 T for $p = 0.08$ and decreases to 40 T for $p = 0.22$. In the UD region, the values of $H_{c2}(T_c)$ estimated from the Nernst signal were larger than $H_{irr}(0)$ estimated in the present study. This is reasonable in that it satisfies the relationship $H_{c2}(0) \geq H_{irr}(0)$. However, in the OD region, the values of $H_{irr}(0)$ were more than twice those of $H_{c2}(T_c)$. For both to be compatible, $H_{c2}(T)$ in the OD region should increase significantly below $T_c$. It should be noted that there are opinions that the finite Nernst effect above $T_c$ does not necessarily originate from superconductivity.[11]

### 4.2 Comparison with Y-123 and Hg-1201

As mentioned in the Introduction, $H_m(0)$ of Y-123[7] and Hg-1201[8] behaves non-monotonically with $p$, exhibiting a minimum for $p \sim 0.12$ and $\sim 0.10$, respectively, as shown in Fig. 9(a). In these studies, $H_m(T)$ was defined as the value of $H$ at which finite resistance appeared (or vortices depinned and moved freely), and $H_m(0)$ was estimated by extrapolating $H_m(T)$ to 0 K, which is essentially the same as the present $H_{irr}(0)$. Notably, in Y-123, it was demonstrated that $H_m(0)$ estimated by this method coincides with $H_{c2}(0)$ determined by thermal conductivity measurements, thus supporting the vortex phase diagram in Fig. 1(b).[7]

The dip structures of $H_m(0)$ in Y-123 and Hg-1201 in specific $p$ regions are considered to originate from the development of a charge order, which competes with and partially suppresses superconductivity. Correspondingly, the signatures of charge order are also found in the $p$-dependence of $T_c$ shown in Fig. 9(b) as shoulder structures near $p = 0.12$ (Y-123) and $p = 0.10$ (Hg-1201). It should be noted that $p$ for Y-123 was estimated using Tallon's formula with corrections for the non-monotonic change in the $c$-axis length,[26] and $p$ for Hg-1201 was estimated from the magnitude of the Seebeck coefficient at room temperature.[27]

In contrast, in the present study of Bi-2212, there was no signature of discontinuities and/or dip structures in either $T_c$ or $H_{irr}(0)$. Because the parabolic $p$-dependence of $T_c$ [Eq. (1)] is assumed, one may overlook the dip in the $T_c$ dome, even if it exists. It is worth noting that $T_c$ of non-substituted and Y-substituted Bi-2212 likely follows the parabolic $p$-dependence,[18,28,29] whereas the $p$-dependence of $T_c$ of Zn-substituted Bi-2212 shows a shoulder structure around $p = 1/8$,[29] similar to Y-123 and Hg-1201. Is it possible that we missed the signature of the charge order because of an incorrect estimate of $p$? At this moment, this possibility is unlikely because we confirmed that in the UD region, both $T_c$ and $H_{irr}(0)$ increased continuously and monotonically by annealing the samples in a more oxidizing atmosphere. In other words, we did not observe any non-monotonic decrease in $T_c$ or $H_{irr}(0)$ under certain annealing conditions. Additionally, in Y-123 and Hg-1201, the charge order is characterized by a lower $H_m(0)$ despite a higher $T_c$, which was not observed in the Bi-2212 samples in this study. At this point, there was no signature in $H_{irr}(0)$ suggesting a developed charge order in Bi-2212, although several studies reported the existence of charge order in Bi-2212.[30-34] This indicates that the influence of the charge order on $H_{irr}(0)$ in Bi-2212 may be less prominent than that in Y-123 and Hg-1201. Another possible reason is that the increase in anisotropy with decreasing $p$ enhances the thermal fluctuations, resulting in suppression of $H_{irr}(0)$, and thus masks the peak of $H_{irr}(0)$ at the UD region.

Finally, we touch upon the anomalous vortex states (second vortex solid phase,[12] unconventional quantum vortex matter state,[13] or anomalous vortex liquid[14]) emerging at low $T$ ($\leq 5$ K) and high $\mu_0 H$ ($\geq 25$ T), which are considered to be associated with the competition between superconductivity and ordered phases. In their vortex phase diagrams, these anomalous vortex states extend to the region well above $H_{irr}(0)$ estimated by extrapolation from the $H_{irr}(T)$ values at relatively high $T$s, that is, the "true" $H_{irr}(0)$ is possibly much larger. Because of the limited $\mu_0 H$ ($< 24$ T) and $T$ ($4.2-15$ K) ranges in the present experiment, we cannot definitively determine whether such an anomalous state exists and whether our results underestimate $H_{irr}(0)$. However, previous studies have shown that anomalous vortex states appear in a limited $p$ range, where a long-range charge order exists. The $H_{irr}(0)$ value evaluated in this study exhibited a monotonic increase from $p = 0.065$ to $0.200$. Accordingly, if charge order occurs in Bi-2212 at a certain $p$, e.g., $p \sim 1/8$, the (true) $p$-dependence of $H_{irr}(0)$ should show a peak at this $p$. This is incompatible with the idea that superconductivity is suppressed by competing order(s). In future, it will be necessary to conduct experiments on UD Bi-2212 using pulsed magnetic fields, as in previous studies, to elucidate the presence of anomalous phases.

## 5. Conclusions

We investigated $H_{irr}(0)$, which is either equal to, or the lower limit of, $H_{c2}(0)$, for Bi-2212 and Bi(Y)-2212 single crystals over a wide range of $p$ values by measuring the MHLs. $H_{irr}(0)$ was evaluated by extrapolating $H_{irr}(T)$ to 0 K using an empirical exponential $T$ dependence. The obtained $H_{irr}(0)$ increased monotonically with $p$ from the UD to the OD, which was distinct from the parabolic $p$-dependence of $T_c$. The present results are qualitatively consistent with $H_{c2}(0)$ from the specific heat study, whereas the opposite is true for the Nernst effect. Moreover, the monotonic $p$-dependence of $H_{irr}(0)$ in Bi-2212 was distinct from that in Y-123 and Hg-1201, showing pronounced dip structures in the UD region. This demonstrates the significant material dependence of the competition between superconductivity and other orders and provides clues for understanding the phase diagrams of high-$T_c$ cuprate superconductors.

**Acknowledgments**

The authors thank Shin-ichi Uchida and Hideo Aoki for their useful discussions and Naoki Shirakawa for the technical support. The main part of this study was performed at High Field Laboratory for Superconducting Materials, Institute for Materials Research, Tohoku University under the GIMRT program (Projects 202112-HMKPA-0025, 202212-HMKPA-0046, and 202312-HMKPA-0047). This study was supported by JSPS-KAKENHI (No. JP19H05823, and JP21H01377), JST-SPRING (No. JPMJSP2151); and JST, the establishment of university fellowships for the creation of science technology innovation (No. JPMJFS2144).

*E-mail: s.ishida@aist.go.jp, nishio@rs.tus.ac.jp

1) M. Tinkham, *Introduction to Superconductivity: Second Edition*, Dover, Mineola, New York (1996).

2) G. Blatter, M. V. Feigel'man, V. B. Geshkenbein, A. I. Larkin and V. M. Vinokur, Rev. Mod. Phys. **66**, 1125 (1994).

3) N. P. Ong and Y. Wang, Physica C **408**, 11 (2004).

4) Y. Ando, G. S. Boebinger, A. Passner, L. F. Schneemeyer, T. Kimura, M. Okuya, S. Watauchi, J. Shimoyama, K. Kishio, K. Tamasaku, N. Ichikawa and S. Uchida, Phys. Rev. B **60**, 12475 (1999).

5) T. Shibauchi, L. Krusin-Elbaum, M. Li, M. P. Maley and P. H. Kes, Phys. Rev. Lett. **86**, 5763 (2001).

6) B. J. Ramshaw, J. Day, B. Vignolle, D. Leboeuf, P. Dosanjh, C. Proust, L. Taillefer, R. Liang, W. N. Hardy and D. A. Bonn, Phys. Rev. B **86**, 174501 (2012).

7) G. Grissonnanche, O. Cyr-Choinière, F. Laliberté, S. René de Cotret, A. Juneau-Fecteau, S. Dufour-Beauséjour, M.-È. Delage, D. LeBoeuf, J. Chang, B. J. Ramshaw, D. A. Bonn, W. N. Hardy, R. Liang, S. Adachi, N. E. Hussey, B. Vignolle, C. Proust, M. Sutherland, S. Krämer, J.-H. Park, D. Graf, N. Doiron-Leyraud and L. Taillefer, Nat. Commun. **5**, 3280 (2014).

8) M. K. Chan, R. D. McDonald, B. J. Ramshaw, J. B. Betts, A. Shekhter, E. D. Bauer and N. Harrison, Proc. Natl. Acad. Sci. USA **117**, 9782 (2020).

9) Y. Wang, S. Ono, Y. Onose, G. Gu, Y. Ando, Y. Tokura, S. Uchida and N. P. Ong, Science **299**, 86 (2003).

10) Y. Wang, L. Li and N. P. Ong, Phys. Rev. B **73**, 024510 (2006).

11) J. Chang, N. Doiron-Leyraud, O. Cyr-Choinière, G. Grissonnanche, F. Laliberté, E. Hassinger, J. P. Reid, R. Daou, S. Pyon, T. Takayama, H. Takagi and L. Taillefer, Nat. Phys. **8**, 751 (2012).

12) F. Yu, M. Hirschberger, T. Loew, G. Li, B. J. Lawson, T. Asaba, J. B. Kemper, T. Liang, J. Porras, G. S. Boebinger, J. Singleton, B. Keimer, L. Li and N. P. Ong, Proc. Natl. Acad. Sci. USA **113**, 12667 (2016).

13) Y.-T. Hsu, M. Hartstein, A. J. Davies, A. J. Hickey, M. K. Chan, J. Porras, T. Loew, S. V. Taylor, H. Liu, A. G. Eaton, M. Le Tacon, H. Zuo, J. Wang, Z. Zhu, G. G. Lonzarich, B. Keimer, N. Harrison and S. E. Sebastian, Proc. Natl. Acad. Sci. USA **118**, e2021216118 (2021).

14) Y. Te Hsu, M. Berben, M. Čulo, S. Adachi, T. Kondo, T. Takeuchi, Y. Wang, S. Wiedmann, S. M. Hayden and N. E. Hussey, Proc. Natl. Acad. Sci. USA **118**, e2016275118 (2021).

15) M. Mittag, M. Rosenberg, D. Peligrad, R. Wernhardt, V. A. M. Brabers, J. H. P. M. Emmen and D. Hu, Supercond. Sci. Technol. **7**, 214 (1994).

16) C. Dewhurst, D. Cardwell, A. Campbell, R. Doyle, G. Balakrishnan and D. M. K. Paul, Phys. Rev. B **53**, 14594 (1996).

17) J. L. Tallon, J. G. Storey, J. R. Cooper and J. W. Loram, Phys. Rev. B **101**, 174512 (2020).

18) M. R. Presland, J. L. Tallon, R. G. Buckley, R. S. Liu and N. E. Flower, Physica C **176**, 95 (1991).

19) S. Hanai, K. Watanabe, H. Oguro, H. Miyazaki, S. Hanai, T. Tosaka and S. Ioka, IEEE Trans. Appl. Supercond. **24**, 4301204 (2014).

20) S. Awaji, K. Watanabe, H. Oguro, H. Miyazaki, S. Hanai, T. Tosaka and S. Ioka, Supercond. Sci. Technol. **30**, 065001 (2017).

21) C. P. Bean, Rev. Mod. Phys. **36**, 31 (1964).

22) W. A. Fietz and W. W. Webb, Phys. Rev. **185**, 862 (1969).

23) T. Matsushita, *Flux Pinning in Superconductors*, Springer Berlin Heidelberg, Berlin, Heidelberg (2014).

24) A. Morello, A. G. M. Jansen, R. S. Gonnelli and S. I. Vedeneev, Phys. Rev. B **61**, 9113 (2000).

25) A. Schilling, R. Jin, J. D. Guo and H. R. Ott, Phys. Rev. Lett. **71**, 1899 (1993).

26) R. Liang, D. A. Bonn and W. N. Hardy, Phys. Rev. B **73**, 180505 (2006).

27) A. Yamamoto, W. Z. Hu and S. Tajima, Phys. Rev. B **63**, 024504 (2000).

28) I. K. Drozdov, I. Pletikosić, C.-K. Kim, K. Fujita, G. D. Gu, J. C. Séamus Davis, P. D. Johnson, I. Božović and T. Valla, Nature Commun. **9**, 5210 (2018).

29) M. Akoshima, T. Noji, Y. Ono and Y. Koike, Phys. Rev. B **57**, 7491 (1998).

30) J. E. Hoffman, E. W. Hudson, K. M. Lang, V. Madhavan, H. Eisaki, S. Uchida and J. C. Davis, Science **295**, 466 (2002).

31) Y. Kohsaka, C. Taylor, K. Fujita, A. Schmidt, C. Lupien, T. Hanaguri, M. Azuma, M. Takano, H. Eisaki, H. Takagi, S. Uchida and J. C. Davis, Science **315**, 1380 (2007).

32) E. H. d. S. Neto, P. Aynajian, A. Frano, R. Comin, E. Schierle, E. Weschke, A. Gyenis, J. Wen, J. Schneeloch, Z. Xu, S. Ono, G. Gu, M. Le Tacon and A. Yazdani, Science **343**, 393 (2014).

33) M. Hashimoto, G. Ghiringhelli, W.-S. Lee, G. Dellea, A. Amorese, C. Mazzoli, K. Kummer,

N. B. Brookes, B. Moritz, Y. Yoshida, H. Eisaki, Z. Hussain, T. P. Devereaux, Z.-X. Shen and L. Braicovich, Phys. Rev. B **89**, 220511(R) (2014).

34) H. Lu, M. Hashimoto, S.-D. Chen, S. Ishida, D. Song, H. Eisaki, A. Nag, M. Garcia-Fernandez, R. Arpaia, G. Ghiringhelli, L. Braicovich, J. Zaanen, B. Moritz, K. Kummer, N. B. Brookes, K.-J. Zhou, Z.-X. Shen, T. P. Devereaux and W.-S. Lee, Phys. Rev. B **106**, 155109 (2022).

**Figure Captions**

**Fig. 1.** (Color online) Schematic illustration of the vortex phase diagrams of type-II superconductors. (a) When thermal fluctuations are negligible (typical superconductors). (b) When thermal fluctuations are strong. (c) When superconducting and thermal fluctuations are strong. In this case, $H_{c2}$ may not be well defined as indicated by the dotted line.

**Fig. 2.** (Color online) $T$ dependence of normalized magnetization ($M$) in FC process for (a) Bi-2212 annealed in $O_2$, (b) in $O_2/N_2$ mixture, and (c) Bi(Y)-2212. (d) Relationship between $T_c$ and $p$ obtained using the empirical parabolic dependence [Eq. (1)].

**Fig. 3.** (Color online) MHLs of Bi(Y)-2212 measured at $T = 55$–$35$ K (a), $30$–$20$ K (b), and $20$–$4.2$ K (c). The dashed rightward (leftward) arrows indicate $H$-increasing (decreasing). (d–f) Enlarged views near $M = 0$ corresponding to (a–c). Solid vertical arrows indicate $H_{irr}$ at each $T$.

**Fig. 4.** (Color online) (a) $J_c - H$ of Bi(Y)-2212 measured at $T = 18$–$4.2$ K. The horizontal dashed lines indicate $J_c$ criteria (1 (pink), 3 (orange), 5 (blue), and 10 (red) $\times 10^6$ A·m$^{-2}$). (b) $F_p - H$ calculated from the data in (a) using Eq. (2). Inverse triangles indicate $F_p^{max}$ at each $T$. (c) $f_p -$ $h^*$ obtained by normalizing the data in (b). (d) $F_p^{max}(T)$ plotted against $H_{max}(T)$ in a double-logarithmic plot. (e) $T$-dependence of $H_{irr}/H_{max}$ determined using different $J_c$ criteria.

**Fig. 5.** (Color online) (a) $H_{irr}(T)$ of Bi(Y)-2212 with $T_c = 61.2$ K ($p = 0.094$). (b) Enlarged view of (a). Red diamonds indicate $H_{irr}$ values estimated from $H_{max}$. Blue circles represent those determined by the $J_c$ criterion of $3 \times 10^6$ A·m$^{-2}$. The dashed line in (b) is the result of fitting using Eq. (6) and the data in the $T$ range of $4.2$–$13$ K.

**Fig. 6.** (Color online) Profiles of $J_c(H)$ [(a1)–(a10)], $F_p(H)$ [(b1)–(b10)], and $f_p(h^*)$ [(c1)–(c10)] for $p$ values of $0.073$ to $0.200$. The $\mu_0 H$ range of the horizontal axis is $0$–$10$ T for upper panels ($p = 0.073$–$0.116$) and $0$–$26$ T for lower panels ($p = 0.121$–$0.200$).

**Fig. 7.** (Color online) (a) $H_{irr}(T)$ for all samples. Circles and triangles indicate the data for Bi-2212 and Bi(Y)-2212, respectively. The solid lines are the results of exponential fitting for $H_{irr}(T)$. (b) $p$-dependence of $H_{irr}(0)$ derived from (a). The blue circles and red triangles represent

$H_{irr}(0)$ for Bi-2212 and Bi(Y)-2212, respectively. The error bars indicate the uncertainty of $H_{irr}/H_{max}$ ($= 5 \pm 0.5$). The dashed line is the $p$-dependence of $T_c$, with their values annotated on the right y-axis.

**Fig. 8.** (Color online) $p$-dependence of $H_{irr}(0)$, $H_{c2}(0)$, and $H_{c2}(T_c)$ for Bi-2212. The blue circles represent $H_{irr}(0)$ evaluated from MHLs in this work. The red diamonds indicate $H_{c2}(0)$ analyzed from the specific heat measurements.[17] The green squares show $H_{c2}(T_c)$ evaluated from the measurements of the Nernst effect.[9]

**Fig. 9.** (Color online) (a) $p$-dependence of $H_{irr}(0)$ for Bi-2212 (blue circles, this work), $H_m(0)$ for Y-123 (red squares),[7] and $H_m(0)$ for Hg-1201 (green diamonds).[8] (b) $p$-dependence of $T_c$ for Bi-2212, Y-123, and Hg-1201. The color code is the same as in (a). Lines are guides to the eye.

**Table**

Table I. List of Bi-2212 and Bi(Y)-2212 samples and corresponding post-annealing conditions, and $p$ estimated from $T_c$.

| Sample number | Post-annealing condition | | | $p$ | $T_c$ (K) |
|---|---|---|---|---|---|
| | Atmosphere | $O_2$ partial pressure (atm) | $T_a$ (°C) | | |
| | Bi(Y)-2212 / $Bi_{2.04}Sr_{1.96}Ca_{1-y}Y_yCu_2O_{8+\delta}$ | | | | |
| #Y1 | $O_2 / N_2$ | $7 \times 10^{-4}$ | 600 | 0.065 | 23.7 |
| #Y1 | $O_2$ | 1 | 800 | 0.073 | 35.6 |
| #Y2 | $O_2$ | 1 | 800 | 0.088 | 54.5 |
| #Y1 | $O_2$ | 1 | 400 | 0.094 | 61.2 |
| #Y2 | $O_2$ | 1 | 400 | 0.116 | 79.6 |
| | Bi-2212 / $Bi_{2.16}Sr_{1.84}CaCu_2O_{8+\delta}$ | | | | |
| #2 | $O_2 / N_2$ | $1 \times 10^{-5}$ | 420 | 0.092 | 53.8 |
| #4 | $O_2 / N_2$ | $1 \times 10^{-5}$ | 630 | 0.105 | 65.1 |
| #2 | $O_2 / N_2$ | $1 \times 10^{-5}$ | 630 | 0.110 | 69.7 |
| #4 | $O_2 / N_2$ | $7 \times 10^{-6}$ | 570 | 0.121 | 76.5 |
| #1 | $O_2 / N_2$ | $8 \times 10^{-6}$ | 450 | 0.136 | 83.4 |
| #2 | $O_2$ | 1 | 800 | 0.160 | 87.6 |
| #1 | $O_2$ | 1 | 600 | 0.178 | 85.4 |
| #3 | $O_2$ | 1 | 500 | 0.189 | 81.8 |
| #3 | $O_2$ | 1 | 400 | 0.200 | 76.1 |

**Fig. 1.**

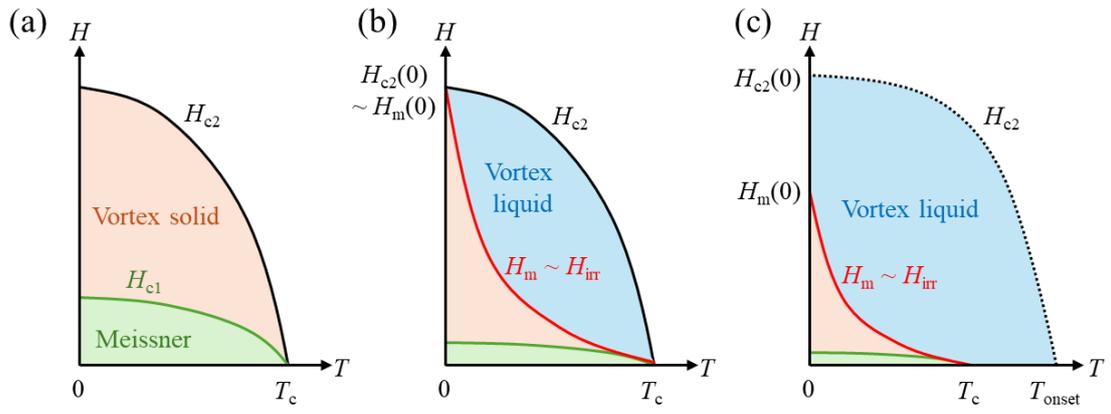

**Fig. 2.**

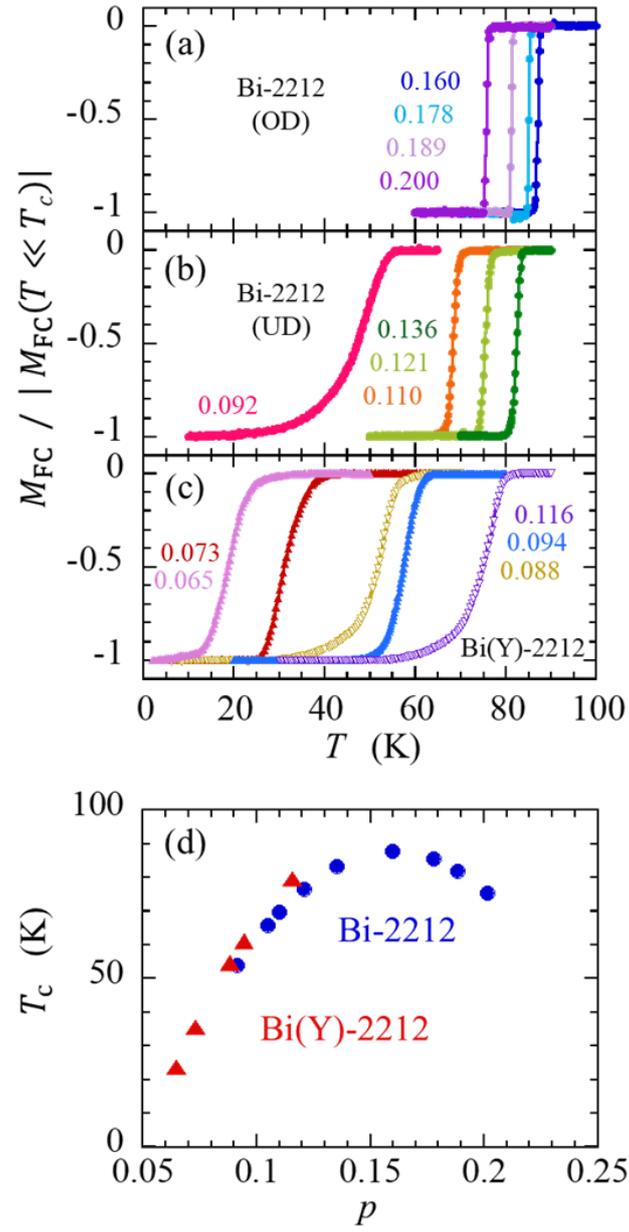

**Fig. 3.**

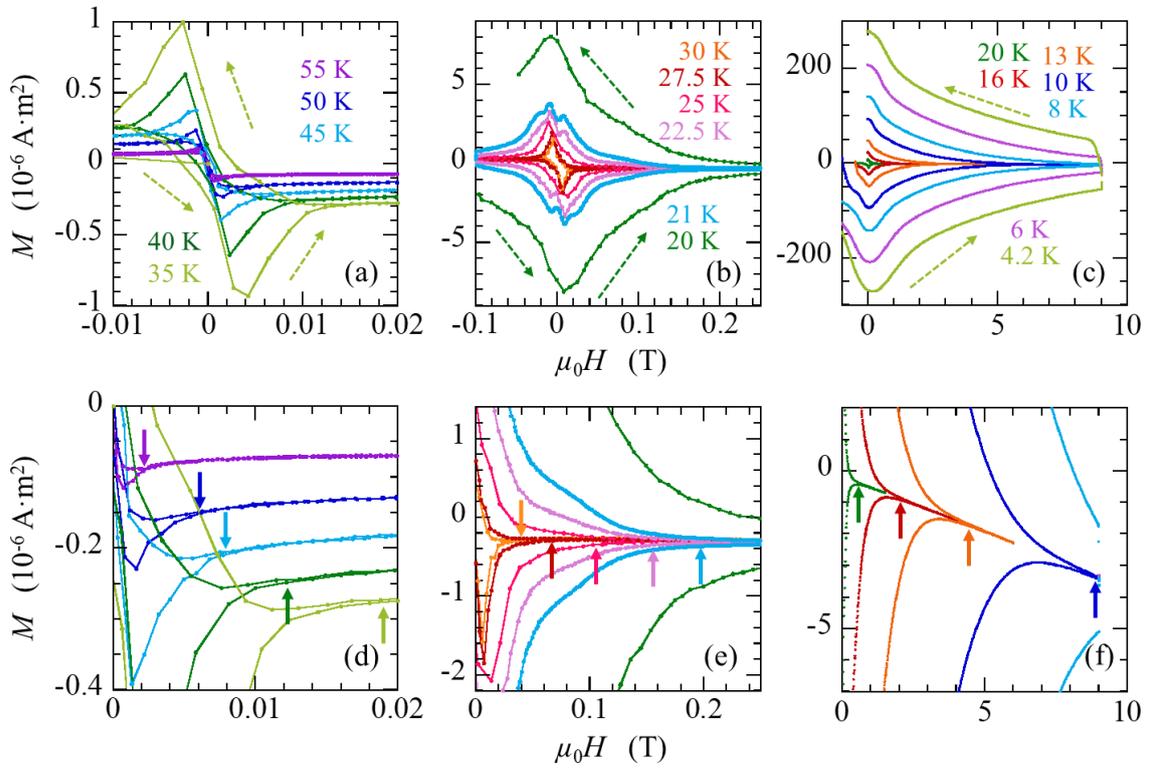

**Fig. 4.**

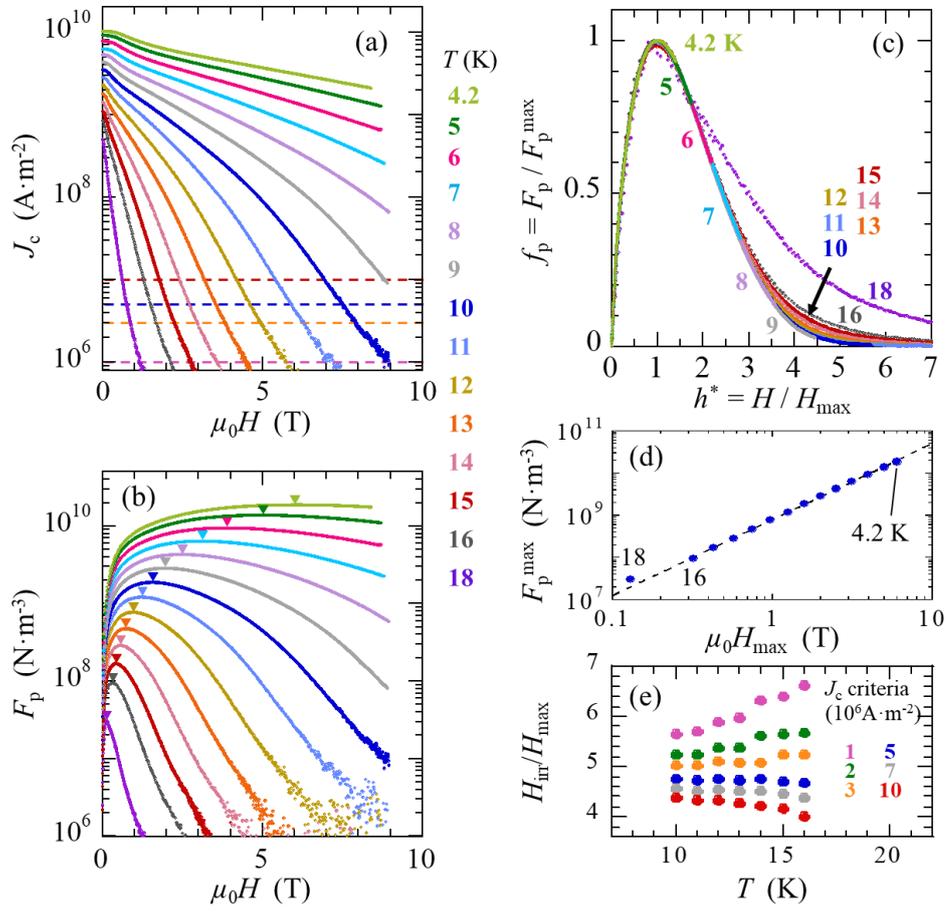

**Fig. 5.**

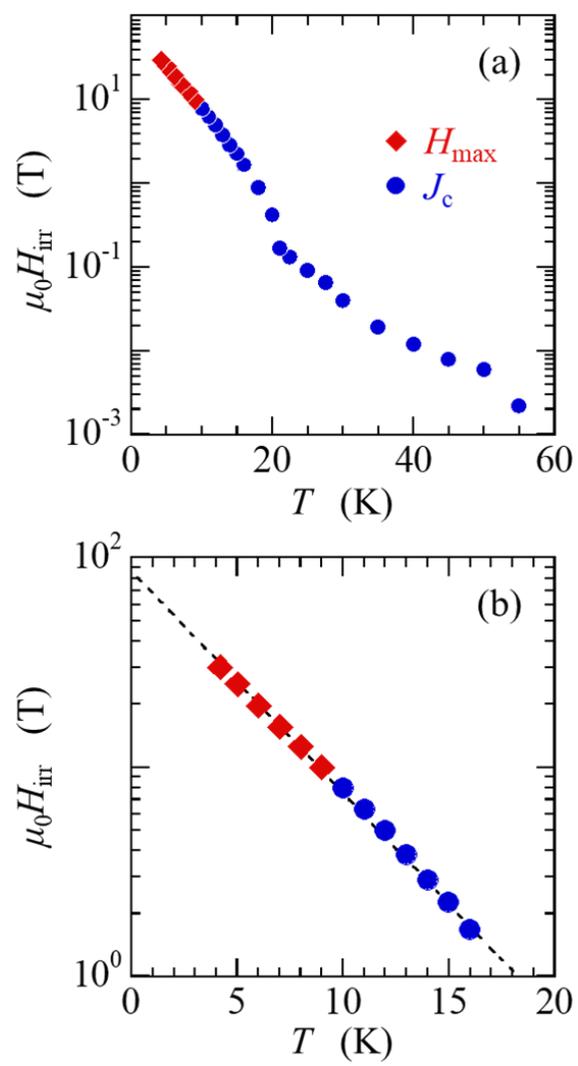

**Fig. 6.**

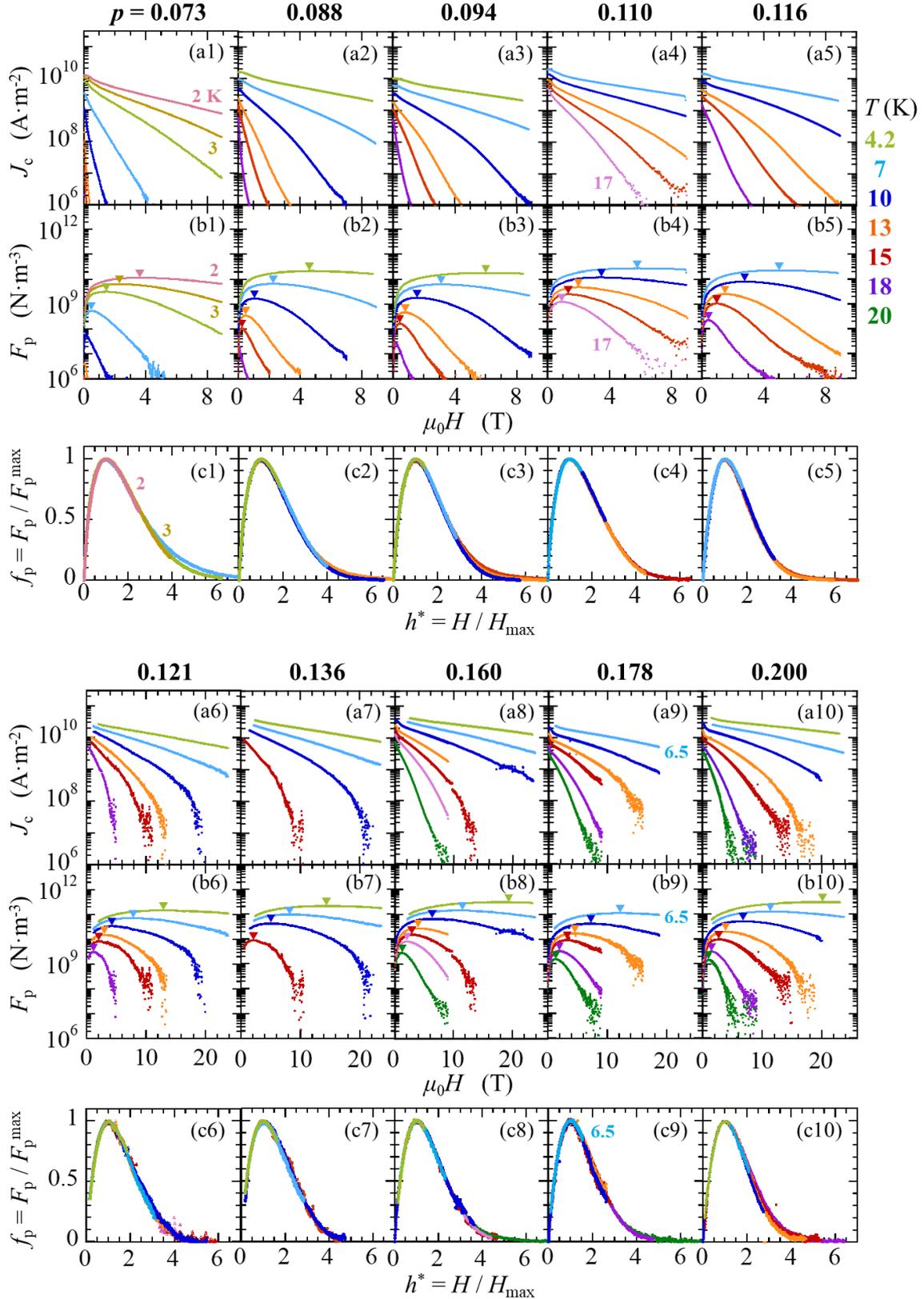

**Fig. 7.**

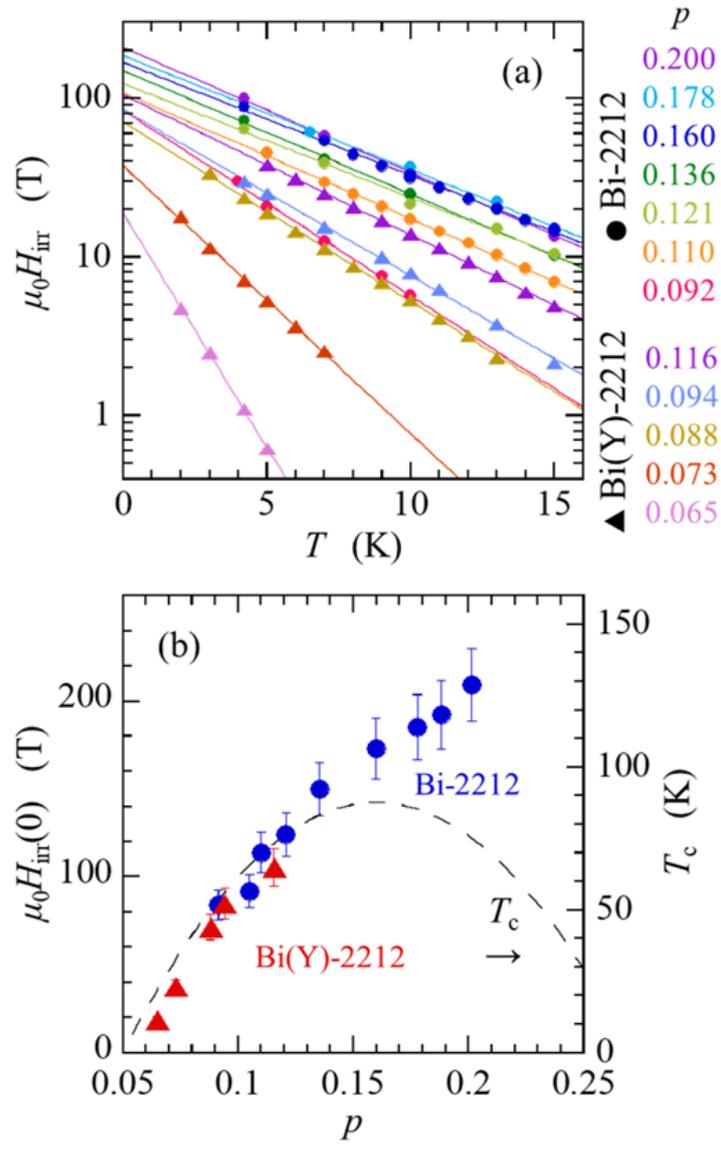

**Fig. 8.**

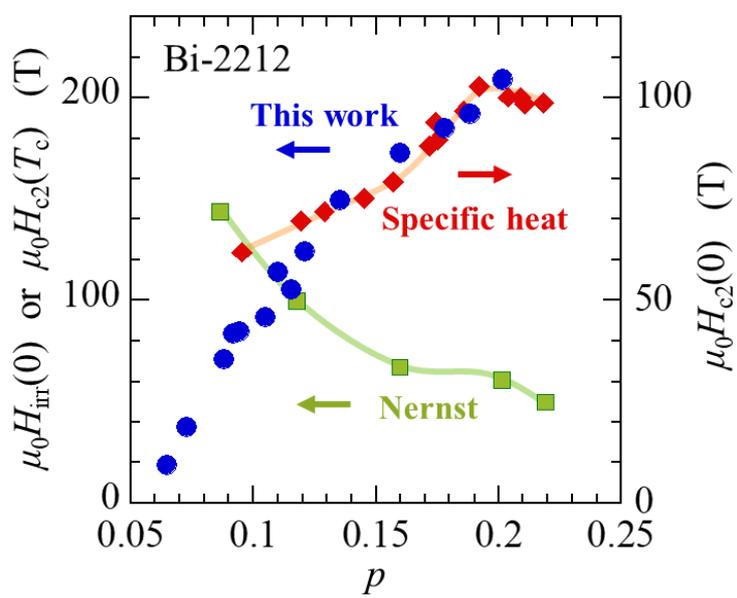

**Fig. 9.**

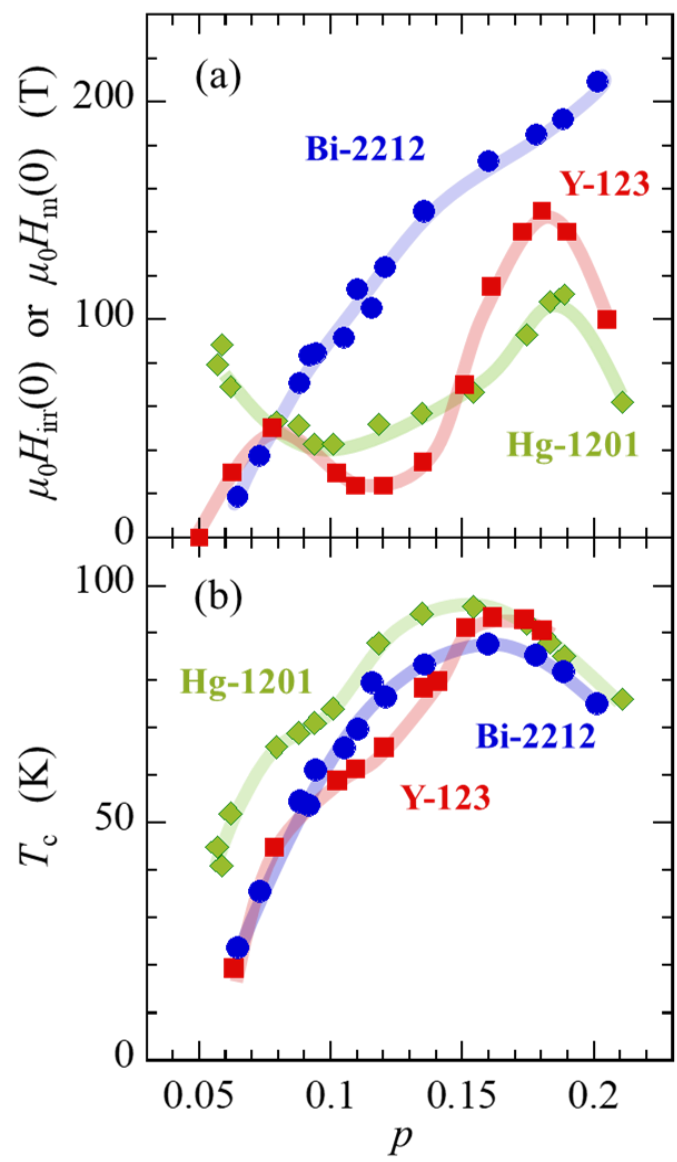